\documentclass[twocolumn,11pt]{article}
\usepackage{amssymb}
\usepackage{amsmath}
\newtheorem{theorem}{Theorem}

\newtheorem{deff}{Definition}

\newcommand{\ignore}[1]{}

\newcommand{\beq}[1]{\begin{equation}}
\newcommand{\enq}[0]{\end{equation}}

\newcommand{\ra}{\rangle}
\newenvironment{proof}{{\noindent{\bf Proof:}}}{$\Box$}

\title{A Simple Proof that 
 Toffoli and Hadamard are Quantum Universal}
\author{Dorit Aharonov\thanks{School of Engineering and Computer Science, The Hebrew University, Jerusalem, Israel, and the Mathematical Sciences Research Institute, Berkeley, California}}
\date{}

\begin{document}
\pagestyle{plain}
\pagenumbering{arabic}

\maketitle
\begin{abstract}
Recently Shi \cite{shi} proved that Toffoli and Hadamard are universal
for quantum computation. This is perhaps the simplest
 universal set 
of gates that one can hope for, conceptually;  It shows that one 
only needs to add 
the Hadamard gate to make a 'classical' set of gates quantum universal.
In this note we give a few lines proof of this fact 
relying on Kitaev's universal set of gates \cite{kitaev}, 
and discuss the meaning of the result.  
\end{abstract}

\section{Introduction}
Quantum computers, believed to be computationally stronger 
than classical devices, 
are constructed from elementary quantum building blocks, 
namely {\it qubits} and {\it gates}. The gates are drawn 
from a {\it universal} set of gates, namely, a set 
which can be used to perform general quantum computation.  
Following the pioneering result of universality of three qubit gates
by Deutsch \cite{deutsch}, 
the question of universality
 has been studied extensively, and a wide array of
sets of gates were proven to be universal
(Just a few examples: \cite{barenco, divincenzo, adleman, lloyd,
 kitaev, aharonov, shor, mor, bacon}.)

 Given the existence of
convenient sets such as the one consisting of all one qubit gates plus the 
gate controlled-NOT \cite{barenco}, 
one might wonder why it is interesting to prove universality of 
many more different sets of gates. 
 The reason
is that different sets of gates are suitable for different tasks.   
For example, to implement a certain algorithm in the laboratory, 
one would use a restricted set which consists of 
gates which are possible to implement in the particular physical realization 
of a quantum computer. However, the theoretical design of
 the algorithm might be easier 
using a completely different universal set. 
Often, one is interested in fault tolerant implementation 
of quantum circuits, in which case one is interested in universal sets 
 consisting of gates which can be implemented 
 fault tolerantly. 
Thus, in quantum computation, we often translate between 
different universal sets; These can be viewed as different {\it programming 
languages} encoding the same algorithm. 
The reason we can interchange 
between different sets without it being too costly both 
of the number of gates and in terms 
of calculating the description of the new circuit, is a deep theorem 
due to Solovay and Kitaev \cite{nielsen} which states that 
 translations between different universal sets is not too costly;    
It causes only a polylogarithmic overhead. 
This fact allows freedom in choosing the building blocks 
suitable for our purposes, be it designing algorithms, lower bounding  
the quantum computational power, fault tolerant constructions, 
or experimental implementations. 

Apart from practical and convenience reasons, there is 
a purely philosophical reason to study different universal sets. 
A fundamental question 
in the theory of quantum computation complexity 
is  where does the quantum computational power
come from. From the physical point of view the question is 
what physical systems are capable of performing truly quantum 
evolutions. There have been several surprising results in this direction.  
The Gottesman-Knill theorem \cite{nielsen} shows that
computation using   
CNOT and Hadamard gates can be simulated 
efficiently by classical computers.   
Valiant \cite{valiant} and Terhal and DiVincenzo 
\cite{terhal} gave classical simulations for 
another restricted set of gates, relating them 
to non interacting fermionic systems. 
On the other hand,  Bacon {\it et. al.} \cite{bacon} showed the surprising 
result that 
the exchange interaction is universal. 
Naturally, we would like to understand 
what sets of gates achieve the full quantum computational power
and when is this power lost. 

One of the most natural sets of gates to consider in this study  
is the set consisting of the Toffoli gate denoted $T$ and the 
Hadamard gate denoted $H$. 
 The question of universality of the set $\{T,H\}$ 
was asked by various researchers, 
and was recently solved by Shi \cite{shi}
to the affirmative.  
The purpose of this note is to give a few lines proof 
of the universality of 
 $\{T,H\}$ based on a simple reduction to a known universal set, 
due to Kitaev \cite{kitaev}.

The fact that $\{T,H\}$ is universal has philosophical interpretations. 
The Toffoli gate $T$
 can perform exactly all {\it classical} reversible computation. 
The result says that Hadamard is all that one needs 
to add to classical computations in order to 
achieve the full quantum computation power; 
It perhaps explains the important role that the Hadamard gate plays in  
quantum algorithms, and can be interpreted as saying that 
Fourier transform is really all there is to quantum computation
on top of classical, since the Hadamard gate is the Fourier transform 
over the group $Z_2$. From a conceptual point of view, this is  
perhaps the simplest and most natural universal set 
of gates that one can hope for. 

In the rest of the note
 we define universality more rigorously,  prove the universality of 
$\{T,H\}$ and conclude with a few remarks.

\section{Preliminaries}
We will use the following notation.
\begin{equation}
X=\left(\begin{array}{cc}0&1\\ 1&0\end{array}\right),H=\frac{1}{\sqrt{2}}\left(\begin{array}{cc}1&1\\ 1&-1\end{array}\right),
\end{equation}
\begin{equation}\nonumber
Z=\left(\begin{array}{cc}1&0\\ 0&-1\end{array}\right),
P(i)=\left(\begin{array}{cc}1&0\\ 0&i\end{array}\right).
\end{equation}
Given a matrix $U$ on $k$ qubits, 
$\Lambda(U)$ is the gate which applies $U$
on the last $k$ qubits conditioned that the first qubit is in the state 
$|1\ra$, and does nothing otherwise. The Toffoli gate
denoted by  $T$ is the gate that applies NOT, or $X$ on the third qubit 
conditioned that the first two qubits are in the state $|1\ra$; 
It can be written as $\Lambda^2[X]$.
$U[j_1,..j_k]$ denotes a gate $U$ operating on qubits 
$j_1,...j_k$. 
We assume that all our quantum circuits use 
gates that operate on a constant number of qubits. 

Bernstein and Vazirani showed that quantum circuits can 
be transformed to circuits that use only 
real matrices \cite{bv}. 
This is done by adding one extra  
qubit to the circuit, the state of which 
indicates whether 
 the system's state is in the real or imaginary part 
of the Hilbert space, and replacing 
each complex gate $U$ operating on $k$ 
qubits by its {\it real version},  
denoted $\tilde{U}$, which operates on the same $k$ qubits plus the extra 
qubit. $\tilde{U}$ is defined by: 
\begin{deff}\label{real}
\begin{eqnarray}\nonumber
&&\tilde{U}|i\ra|0\ra= [Re (U)|i\ra]|0\ra+[Im(U)|i\ra]|1\ra\\\nonumber
&&\tilde{U}|i\ra|1\ra=-[Im(U)|i\ra]|0\ra+[Re(U)|i\ra]|1\ra.
\end{eqnarray}
\end{deff}
Here $Re(U),Im(U)$ means the real and the imaginary part of the matrix $U$, 
respectively.  
The new circuit computes the same function with the overhead of one qubit.

\section{Universality}
There have been several notions of Universality used in the literature. 
The strongest sense of universality of a set of gates $S$ is as follows: 
\begin{deff}{\bf (Strict Universality)} 
A set of quantum gates $S$ is said to be strictly universal  
if there exists a constant $n_0$ such that for any $n\ge n_0$,  
the subgroup generated by $S$ is dense in $SU(2^n)$, the group 
of unitary matrices with determinant $1$ operating on $n$ qubits. 
\end{deff}

This means that  any unitary matrix on $n$ qubits
 can be approximated (in the standard operator norm induced by the 
$l_2$ norm)
to within arbitrary accuracy by applying 
a sequence of gates from $S$. 
The determinant is taken to be $1$ since an 
 overall phase can be added to the gates without changing anything.  
Typically $n_0$ is a small number; $2$ or $3$.

Note that the above definition does not require anything regarding 
the {\it rate} of approximation, which in principle can be 
arbitrarily slow;   
Fortunately, the Solovay-Kitaev theorem \cite{nielsen, kitaev} 
guarantees that fast approximation is implied by the definition. 
The theorem states that 
for any fixed $k\ge n_0$, the number of gates from a universal set required 
to approximate a matrix $U$ on $k$ qubits to within $\epsilon$ grows 
only like $polylog(1/\epsilon)$.  Moreover,  
the calculation of the description of the sequence of gates 
approximating $U$ can also be done efficiently.

Most of the universal sets of gates that have appeared in the
literature are of the strictly universal type.
Such is for example Kitaev's set of gates: 
\begin{theorem} (Kitaev \cite{kitaev})\label{kituniv}
The set  $\{\Lambda(P(i)),H\}$ is strictly universal with $n_0=2$.
\end{theorem}

In fact, such a strong notion of universality is 
unnecessary, and often, weaker notions of universality are used.   
For example, universality which allows  
using ancilla states, as is done in Shi \cite{shi};  
In this case, instead of approximating $U$ on the input state $|\xi\ra$ 
 one attempts at approximating 
$U\otimes I$ on the state $|\xi\ra\otimes |\phi\ra$ where $|\phi\ra$
is an ancilla state that can be generated using the same set of gates. 
Another relaxation of the definition is used  
in the context of certain fault tolerant constructions \cite{bacon},
where one is interested in {\it encoded universality}, 
namely generating all unitary matrices on some part 
of the Hilbert space. The Solovay-Kitaev theorem does not hold automatically 
for these weaker definitions of
universality, and one should be careful to check
on a case by case basis  
that fast approximation indeed holds. 

We generalize these relaxations in a definition of {\it computational
 universality}. This definition captures essentially 
the meaning of the notion of universality, 
namely, that the set of gates can be used 
to perform general quantum computation, without too much overhead.

\begin{deff}{\bf (Computational Universality)} 
A set of quantum gates $C$ 
is said to be Computationally Universal  
if it can be used to simulate to within $\epsilon$ error any 
quantum circuit which uses $n$ qubits and $t$ gates from 
 a strictly universal set
with only polylogarithmic overhead in $(n,t,1/\epsilon)$.   
\end{deff}

Strict universality obviously implies computational universality. 
Translations between universal sets satisfying either one of the
definitions are polylogarithmic,  and so for all purposes, 
it is sufficient to prove computational universality
for a set of gates.


\section{Proof}
Here we are interested in the set $\{T,H\}$. 
We cannot hope to prove strict universality 
since the set consists only of real gates and cannot approximate 
complex unitary matrices. 
We show 

\begin{theorem}
The set $\{T,H\}$ is computationally universal. 
\end{theorem}

\begin{proof}
Let $Q$ be a circuit that uses $t$ gates from $S=\{\Lambda(P(i)),H\}$. 
$S$ is a strictly universal set of gates by theorem \ref{kituniv}. 
We replace gates from $S$ by gates from our set $\{T,H\}$. 
 $H$ is already in our set, so we only need to deal with 
 $\Lambda(P(i))$, which we convert to its real version 
using definition \ref{real}. We find that 
 $\widetilde{\Lambda(P(i))}=\Lambda^2[XZ]$ by checking its operation on the $8$ 
basis states.
Since $XZ=XHXH$, this matrix is exactly a product of four gates from our set: 
$T(1,2,3)H(3)T(1,2,3)H(3)$. 
We have simulated $Q$ with at most $4t$ gates, 
and one additional qubit. 
\end{proof}

\section{Concluding Remarks}

We have given a simple proof
 that the set $\{T,H\}$ is computationally universal.  
For stronger results regarding this set, showing that 
it generates a dense subgroup in the group of orthogonal matrices, 
see \cite{shi}. 

We  remark that an inherent disadvantage of 
using real matrices as the universal set of gates 
is that the additional extra qubit is used in many gates, 
which prevents full parallelization of the computation. 
One might worry that for this reason this set cannot be useful for 
fault tolerance against decoherence in the wires, 
which requires parallelism \cite{aharonov3}. 
However this is not true;  
Error correction can still be performed in parallel, since the gates used in 
error corrections, namely 
Hadamard and classical gates, are real and do not require 
the additional qubit, and so $\{T,H\}$ can be used for 
fault tolerant purposes.

\section{Acknowledgements}
I am grateful to Ashwin Nayak and Yaoyun Shi
for helpful discussions.


\begin{thebibliography}{99}
\bibitem{adleman}
L. Adleman, J. Demarrais and M. D. Huang,
Quantum Computability,
{\em SIAM Journal of Computation}
{\bf 26} 5 pp 1524--1540 October, 1997  
\bibitem{aharonov} D. Aharonov, M. Ben-Or, Fault Tolerant Quantum computation 
with Constant Error Rate, quant-ph/9906192
\bibitem{aharonov3}
 D. Aharonov, M. Ben-Or, Polynomial Simulations of Decohered Quantum Computers, in FOCS pp 46--55, 1996
\bibitem{bacon} D. Bacon, J. Kempe, D.P. DiVincenzo, D.A. Lidar, K.B. Whaley,
Encoded Universality in Physical Implementations of a Quantum Computer,
in {\it  Proceedings of the International Conference on Experimental
     Implementation of Quantum Computation}, Sydney, Australia (IQC 01)
\bibitem{barenco}
A. Barenco, C. H. Bennett, R. Cleve, D. P. DiVincenzo, N. Margolus,
P. Shor, T. Sleator, J. Smolin and H. Weinfurter, 
Elementary gates for quantum computation,
{\em Phys. Rev. A} {\bf 52}, 3457--3467, 1995
\bibitem{bv} E. Bernstein and U. Vazirani, Quantum Complexity Theory, Siam J. of Comp. 26(5):1411-1473, 1997, quant-ph/9701001 
\bibitem{mor}  O. Boykin, T. Mor, M. Pulver, V. Roychowdhury, F. Vatan, 
On Universal and Fault-Tolerant Quantum Computing, quant-ph/9906054
\bibitem{deutsch}
D. Deutsch,
Quantum computational networks,
In {\em Proc. Roy. Soc. Lond.} A {\bf 425} 73-90, 1989
\bibitem{deutsch1}
D. Deutsch, A.  Barenco and A. Ekert,  
Universality in quantum computation,
In {\em  Proc. R. Soc. Lond.} A {\bf 449} 669-677, 1995
\bibitem{divincenzo}
D. P. DiVincenzo,
Two-bit gates are universal for quantum computation,
{\em Phys. Rev. A} {\bf 51} 1015-1022 1995
\bibitem{kitaev} A. Yu. Kitaev, Quantum Computations: Algorithms and Error correction, Russian Math. Surveys {\bf 52} no. 6 1191-1249 (1997)
\bibitem{knill}
E. Knill, R. Laflamme, and W. Zurek,
\newblock Resilient quantum computation,
 {\em Science}, vol 279, p.342, 1998. 
\bibitem{lloyd}
S. Lloyd 
Almost any quantum logic gate is universal,
Phys. Rev. Lett. {\bf 75}, 346-349, 1995
\bibitem{nielsen}M. A. Nielsen and I. Chuang,
                                    Quantum Computation and Information,
                                   Cambridge
                                    University Press, 2000
\bibitem{shi} Y. Shi, Both Toffoli and controlled-Not need 
little help to do universal quantum computation, quant-ph/0205115
\bibitem{shor} P. Shor, Fault tolerant quantum computation, 
In FOCS 56-65, 1996.  quant-ph/9605011.
\bibitem{terhal} B. M. Terhal, D. P. DiVincenzo,
Classical simulation of noninteracting-fermion quantum circuits,
Phys. Rev. A 65, 032325/1-10 (2002)
\bibitem{valiant} L. Valiant, Quantum computers that can be Simulated classically in polynomial time. STOC 2001, also http://www.deas.harvard.edu/$\tilde{}$ valiant
\end{thebibliography}
\end{document}